\documentclass[10pt,a4paper]{article}

\usepackage{amsmath}
\usepackage{graphicx}
\usepackage{amsfonts}
\usepackage{amssymb}
\usepackage{subfigure}
\usepackage{float}
\usepackage{color}

\begin{document}

\renewcommand\floatpagefraction{.8}
\renewcommand\topfraction{.9}
\renewcommand\bottomfraction{.9}
\renewcommand\textfraction{.1}

\noindent\LARGE{\textbf{Resolving structural modifications of colloidal glasses by combining x-ray scattering and rheology}}
\vspace{0.6cm}

\noindent\large{D. Denisov,$^{1*}$
M. T. Dang,\textit{$^1$}
B. Struth,\textit{$^2$}
G. Wegdam,\textit{$^1$}
and P. Schall\textit{$^{1}$}}

\vspace{0.5cm}
\noindent{\textit{$^{1}$~Van der Waals-Zeeman Institute, University of Amsterdam, The Netherlands.}}

\noindent{\textit{$^{2}$~Deutsches Elektronen-Synchrotron, Hamburg, Germany. }}

\noindent{Correspondence and requests for materials should be addressed to D. Denisov (d.denisov@uva.nl)}
\vspace{0.5cm}

\noindent\textbf{Glasses have liquid-like structure, but exhibit solid-like properties. A central question concerns the relation between the structure and mechanical properties of glasses, but structural changes remain difficult to resolve. We use a novel combination of rheology and x-ray scattering to resolve structural changes in colloidal glasses and link them directly to their mechanical behavior. By combining stress and structure factor measurements, we resolve shear induced changes in the nearest neighbor configuration as a function of applied stress, allowing us to elucidate the structural origin of the genuine shear banding transition of glasses.}
\section*{}

A great challenge in the understanding of glasses is the relation between their structure and mechanical properties. When a liquid cools down and arrests in a glass, the liquid structure remains virtually unchanged, while the viscosity and structural relaxation time increase rapidly by many orders of magnitude~\cite{Ediger1996}. Structural changes are believed to be central to the flow and relaxation of glasses, but it is difficult to measure them directly.

Nevertheless, structural parameters are crucial ingredients of plasticity theories of glasses~\cite{Langer,Bouchbinder,Spaepen_77,Argon_79}. The description of their flow, time-dependent elasticity and aging typically relies on structural order parameters that vary as a function of applied shear and time. Interpretation and measurement of such structural order parameters, however, remain challenging: structural changes are much more subtle than those occurring in crystals, and because of the nature of the amorphous structure, it is difficult to define an "ideal" amorphous state~\cite{Kurchan2007}. Hence, the relation between structure and mechanical properties remains poorly understood.

Soft matter has emerged as an important laboratory for exploring the glass transition and structural arrest. A variety of systems including polymer melts, colloidal suspensions, foams, and granular materials show dynamical transitions from a
fluid-like state to an arrested solid-like state with no obvious structural change. Hard sphere systems have played a crucial role as model systems to obtain insight into the arrest and flow of glasses~\cite{ColloidalGlasses,Pusey_2008}. Their structure and dynamics resemble closely that of simple atomic liquids, with a sudden dynamic slow-down occurring at particle volume fractions $\phi_g = 0.58$~\cite{vanMegen1998}, the colloidal glass transition. In these systems, the dynamic arrest has been attributed to the caging by nearest neighbors. Hence, the structure of nearest neighbor configurations should play a crucial role in the understanding of the mechanical properties of glasses.

An important property of amorphous materials is their shear banding instability~\cite{Spaepen_77,Argon_79,SB,Varnik_2003,Shi_2007}. When the applied shear rate becomes of the order of the inverse structural relaxation time, the glass develops shear bands that flow at a much increased rate~\cite{Spaepen_77,Argon_79,chikkadi_schall11,spaepen_75}.
%Shear banding can occur when the steady-state flow stress as a function of strain rate has a negative slope \cite{Fielding2007}. While in anisotropic particle systems such as colloidal rods, the reason for this can be a flow-induced structural phase transition~\cite{1997-Dhont}, in simple liquids or glasses, changes are much more subtle and difficult to measure~\cite{besseling_2010}.
Recent direct imaging of the microscopic strain field~\cite{chikkadi_schall11} showed a characteristic symmetry change from solid to liquid, signaling the fundamental nature of this transition; however, changes in the structure factor could not be measured, and the structural origin of shear banding remained unresolved.

Here, we use a novel combination of x-ray scattering and rheology to link the visco-elastic properties of colloidal glasses directly to their structure. We identify structural changes in the nearest-neighbor configuration of the particles while simultaneously measuring the applied stress and strain with rheology. This allows us to elucidate the relation between structure and mechanical properties of glasses: X-ray scattering produces averages over large sample volumes, enabling us to resolve the small, but distinct structural modification of the glass in response to the applied shear. We probe the glass over a wide range of shear rates and relate shear-induced viscosity changes directly to changes in the structure. This allows us to reveal the structural transition as the glass undergoes the genuine shear banding instability.

\section*{Results}

The new combination of shear and structure measurement is achieved by placing a modified rheometer directly in the beam path of the synchrotron (see Fig. 1a). This setup allows us to measure the stress as a function of strain while simultaneously following the suspension structure with x-ray scattering . The colloidal glass consists of silica particles with a hard-sphere diameter of 50 nm and a polydispersity of $10\%$, suspended in water. The particles interact with a screened electrostatic interaction with a screening length of $\sim 2.7$ nm (see Methods). Samples prepared at different volume fractions below, at, and above $\phi_g$ allow us to elucidate shear-induced structural changes across the colloidal glass transition. We investigate changes in the nearest neighbor configuration of the particles by measuring changes in the first peak of the angle-averaged structure factor $S(q)$. Clear changes can be detected in the position $q_1$, the half-width $w_1$, and the intensity $S_1$ of the first peak; this is demonstrated in Fig. 1b and 1c where we plot $q_1$ and $w_1$, together with the applied shear stress during shear ($\dot{\gamma}=10^{-1}$ s$^{-1}$) and relaxation for $\phi = \phi_g$. We have used Gaussian fitting of the first peak of $S(q)$ to determine precise changes in the peak position and width with sub-pixel accuracy.
Both show that the structure changes immediately with the application of stress and relaxes back with a characteristic relaxation time $t_s$ after we stop the shear. We determine $t_s = 35 $ s from the exponential fit $q_1 \sim \exp(-t/ t_s)$ (pink dashed line in Fig. 1b). This relaxation time is a factor of $2 \cdot 10^6$ larger than the Brownian time $t_B = 1.6 \cdot 10^{-5}$ s estimated for dilute suspensions, indicating that the suspension is deep in the glass state. We note that in addition, a second much longer time scale $t_{s2}\gtrsim 1000 $ s dominates the behavior of the peak width, and is also apparent in the peak position. We associate this longer time scale with the aging of the sample, a typical property of the glassy state.

\begin{figure}
\centering
\subfigure
{\includegraphics[width=0.5\columnwidth]{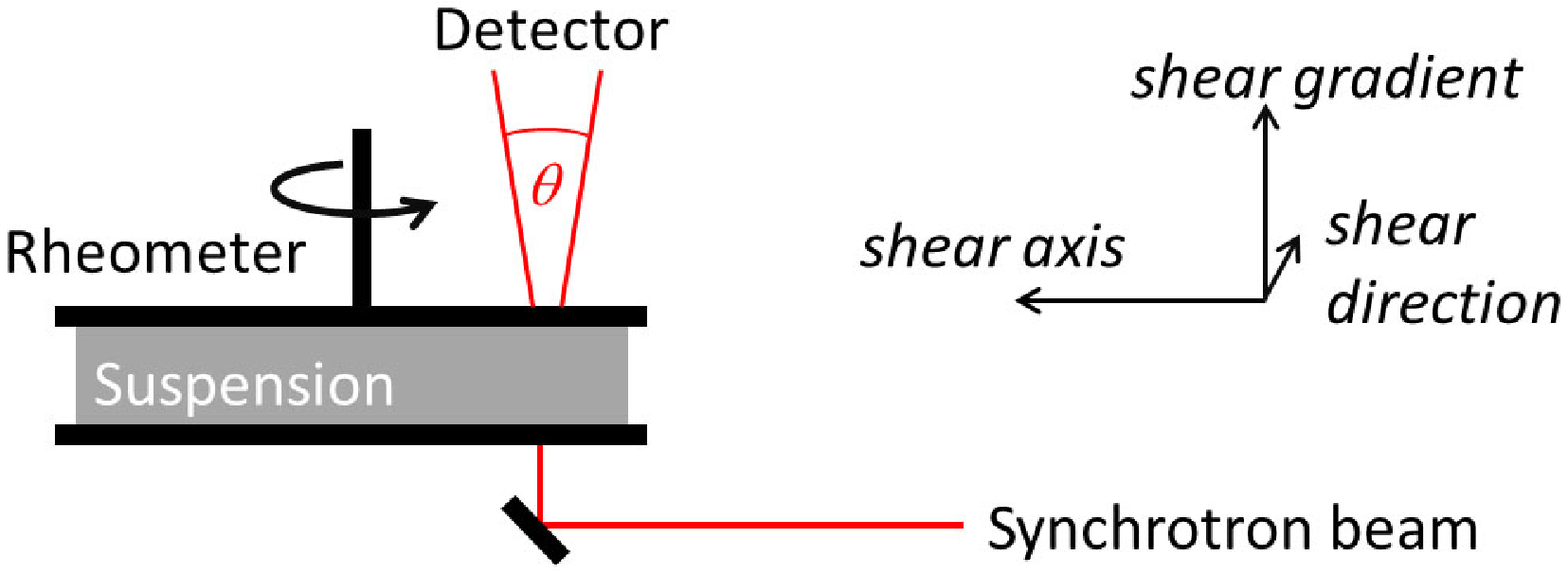}
\label{fig:Setup}
}
\begin{picture}(0,0)(0,0)
\put(-260,00){(a)}
\end{picture}
\subfigure
{\includegraphics[width=0.5\columnwidth]{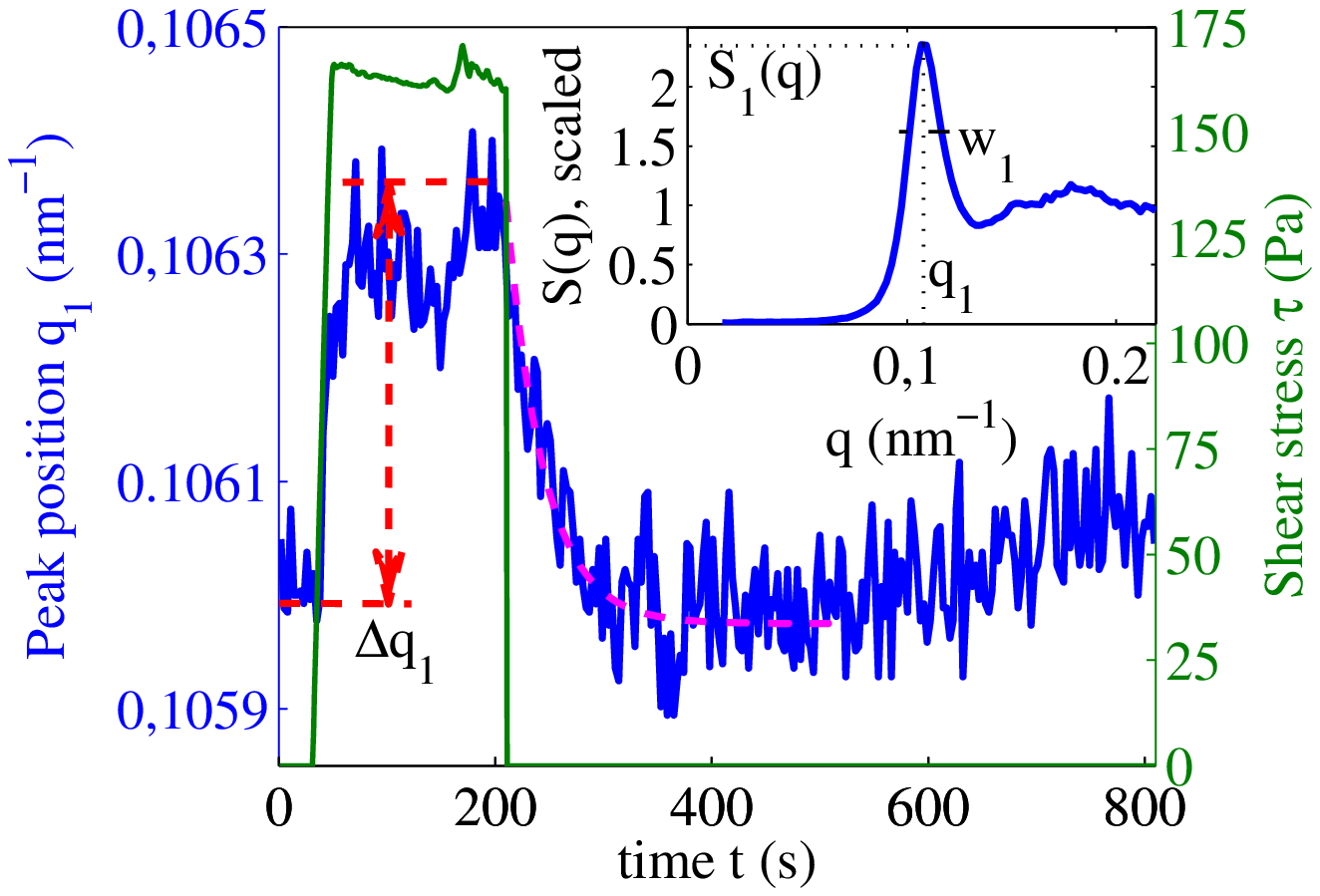}
\label{fig:Diameter_55nm_vf30_g05_01}
}
\begin{picture}(0,0)(0,0)
\put(-260,00){(b)}
\end{picture}
\subfigure
{\includegraphics[width=0.5\columnwidth]{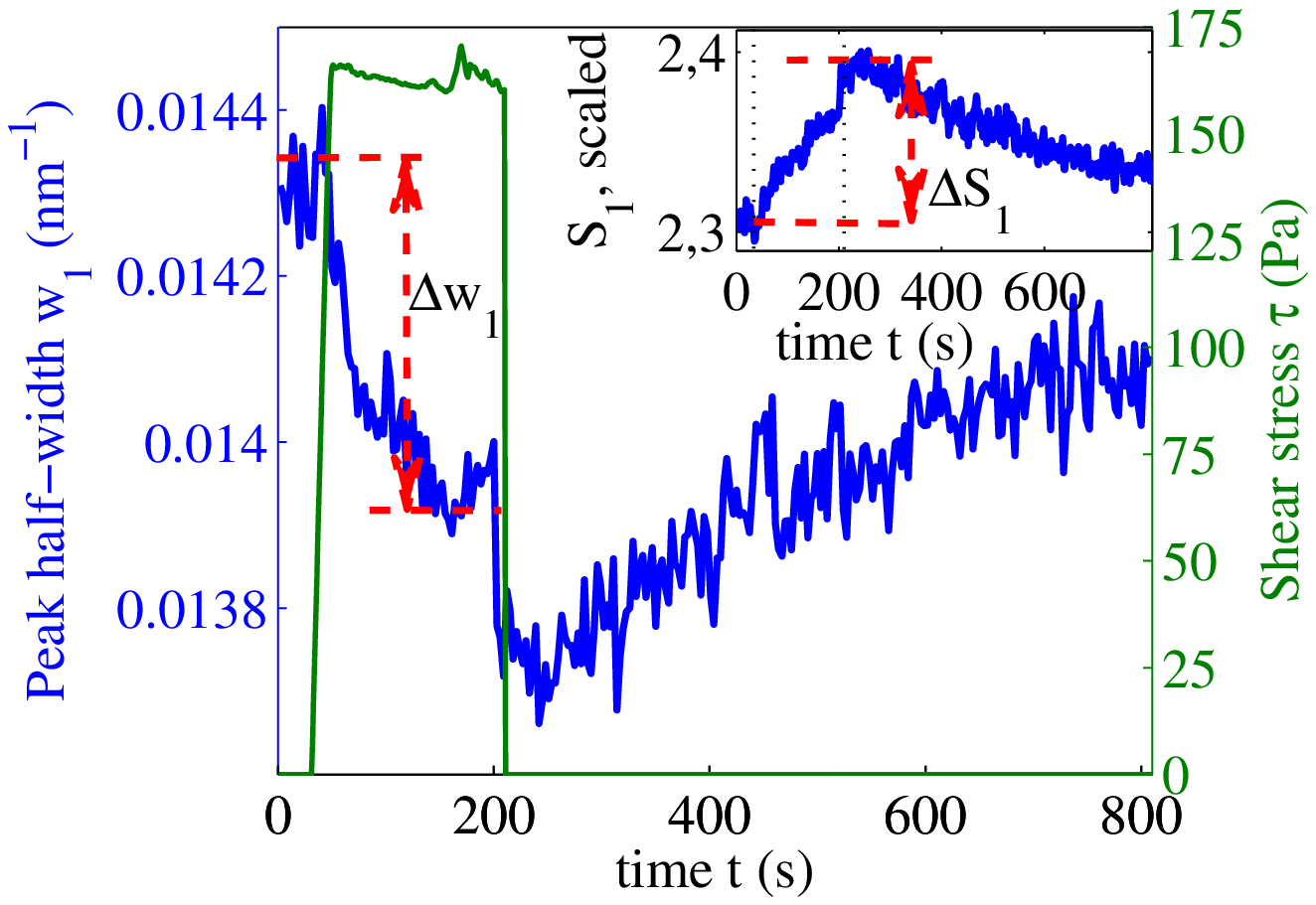}
\label{fig:HalfWidth_55nm_vf30_g05_01}
}
\begin{picture}(0,0)(0,0)
\put(-260,00){(c)}
\end{picture}
\caption{
{\bf Simultaneous rheology and x-ray scattering measurement.}
(a) Schematic of the experimental setup. The synchrotron beam passes through the sheared suspension while the rheometer measures the stress-strain relation. Inset illustrates the shear direction with respect to the incident beam direction.
(b) Nearest-neighbour peak position of the structure factor (left axis, blue) together with the applied shear stress (right axis, green) during shear and relaxation. Data shows the volume fraction $\phi = 58\%$ at the shear rate $\dot\gamma=10^{-1}$ s$^{-1}$. Inset shows the structure factor $S(q)$.
(c) Half-width of the first peak of the structure factor (left axis, blue) and the applied shear stress (right axis, green) during the same experiment. Inset shows the corresponding peak height $S_1$.
\label{fig:DHw_55nm_vf30_g05_01}}
\end{figure}

How does the structure change during the applied shear? The observed increase of $q_1$ indicates that under shear, particles pack more densely along the shear plane; as a result, nearest neighbor correlations grow, as demonstrated by the decrease of the peak width and simultaneous increase of the peak height in Fig. 1c. Assuming constant particle density, we conclude that the vertical separation of the particles must increase, which makes them flow past each other more easily, in agreement with measurements in sheared colloidal crystals~\cite{Cohen2006}. Note that $w_1$ and $S_1$ seem to approach a steady state more slowly than $q_1$, indicating that the ordering of the system happens on a longer time scale.
Thus, Fig. 1 gives direct evidence of small structural changes facilitating the shear flow.

Interestingly, with increasing volume fraction, these isotropic changes diminish (see Table \ref{tab:phi}), reflecting the increasing frustration of the particles. In this case, the structure becomes anisotropic - the intensity distribution $S(q_1)$ along and perpendicular to the shear direction become different, as shown by plotting the structure factor $S_1(\alpha) = S(|q|=q_1,\alpha)$ as a function of angle in Fig. 2a;
%its orientation with respect to the shear direction remains essentially constant during shear and relaxation \cite{intensity_correction}.
These changes, however, are small and dominated by fluctuations. To elucidate the underlying structure and symmetry, we determine the angular correlation function of the fluctuations of $S_1$,

\begin{equation}
C(\beta)=\frac{\int_0^{2\pi}(S_1(\alpha+\beta)-<S_1(\alpha)>)(S_1(\alpha)-<S_1(\alpha)>)d\alpha}
{\int_0^{2\pi}(S_1(\alpha)-<S_1(\alpha)>)^2 d\alpha},
\end{equation}
where $\beta$ is the correlation angle. This allows us to reveal the underlying shear-induced symmetry clearly: distinct two-fold (p-wave) symmetry is observed under applied shear (Fig. 2b, inset). This two-fold symmetry indicates a preferential ordering of particles along planes perpendicular to the shear direction. The ordering follows immediately the applied shear, as is obvious by comparing the time evolution of $C(\beta=\pi)$ and the shear stress in Fig. 2b.

\begin{figure}
\centering
\subfigure
{\includegraphics[width=0.44\columnwidth]{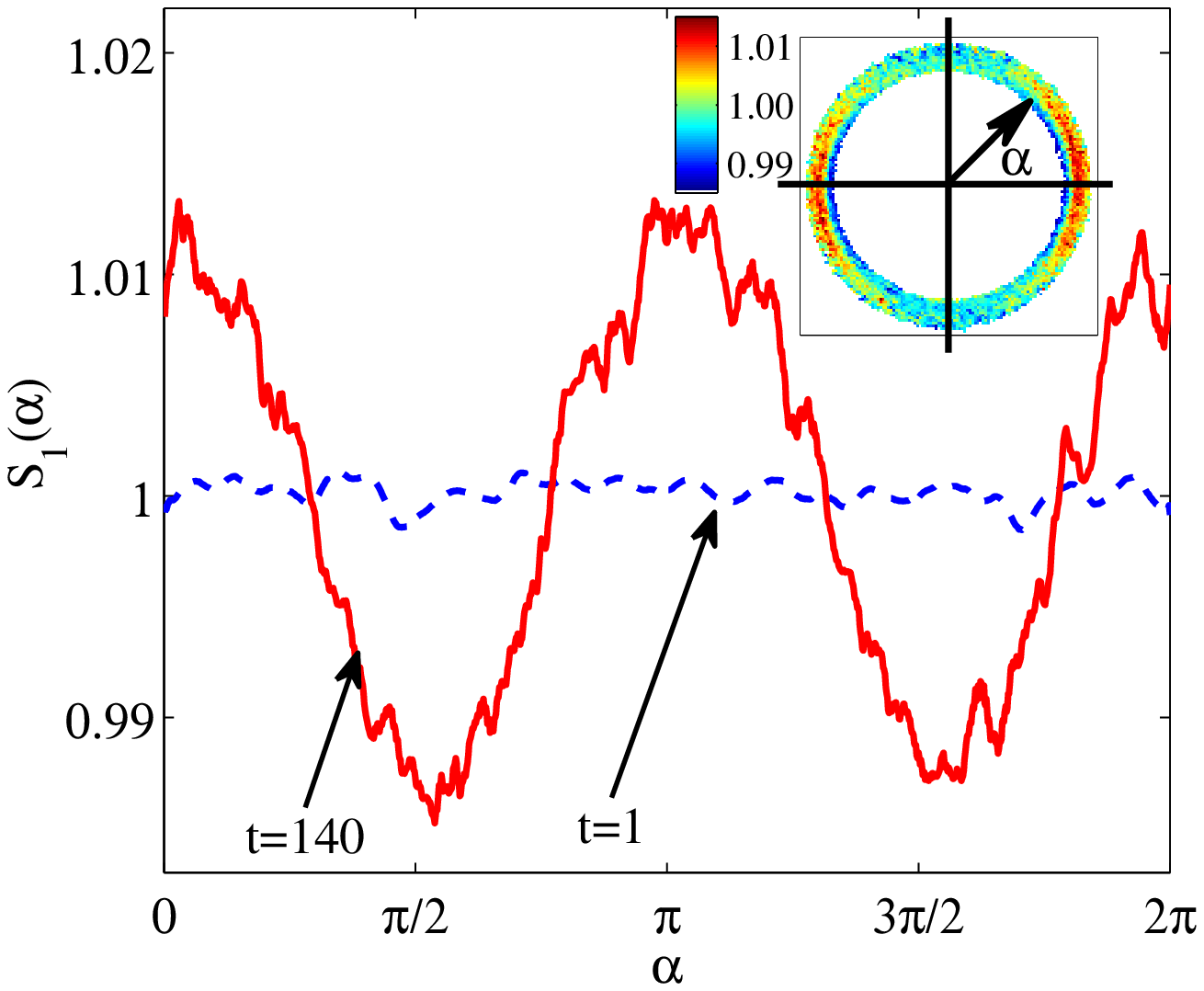}
\label{fig:intensity}
}
\begin{picture}(0,0)(0,0)
\put(-200,0){(a)}
\end{picture}
\subfigure
{\includegraphics[width=0.45\columnwidth]{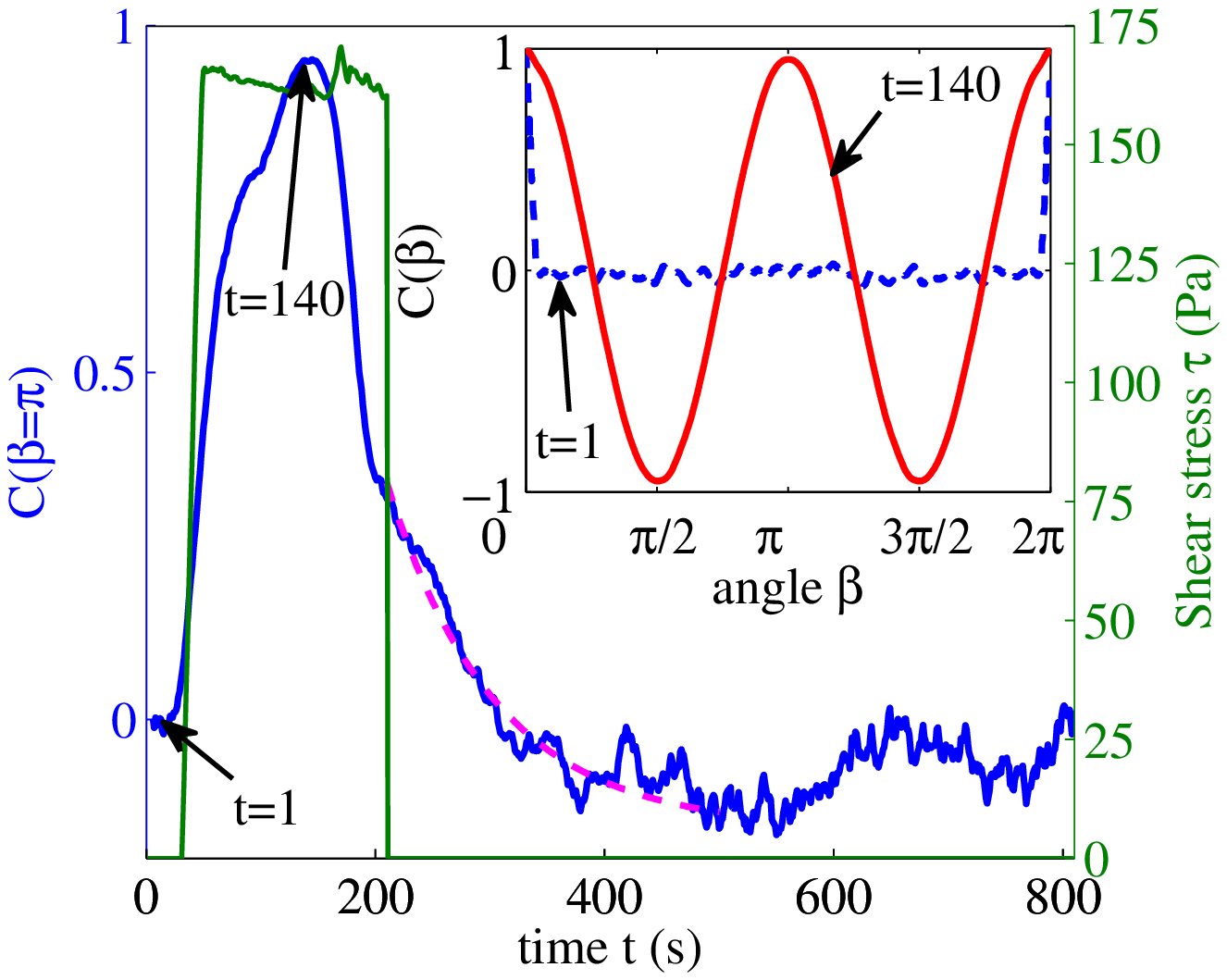}
\label{fig:correlation}
}
\begin{picture}(0,0)(0,0)
\put(-200,0){(b)}
\end{picture}
\caption{
{\bf Angle-resolved structure factor.}
(a) Nearest-neighbour peak value of the structure factor as a function of angle, before application of shear ($t=1$ s, blue dashed line) and during the shear ($t=140$ s, red solid line). Data shows the volume fraction $\phi = 58\%$. Inset shows the intensity distribution along the first ring for $t=140$ s, normalized to the intensity for $t=1$ s. Horizontal axis: shear direction, vertical axis: shear axis.
(b) Time evolution of the symmetry of the structure factor, as measured by the angular correlation function, $C(\beta=\pi,t)$. The correlation function $C(\beta)$ itself is shown in the inset. Distinct two-fold symmetry (solid line) is observed during the shear.}
\label{fig:secondpeak_t}
\end{figure}

\begin{table}[h]
\centering
\begin{tabular}{l  c c c c}
\hline
\hline
$\phi$, $\%$ & $q_1$, nm$^{-1}$ & d, nm & $w_1$, nm$^{-1}$ & $\Delta q_1/q_1\cdot10^2$ \\
\hline
34.5 & 0.089 & 70.62 & 0.0152 & 0.9006 \\

58 & 0.106 & 59.41 & 0.0143 & 0.3138 \\

63.5 & 0.113 & 55.60 & 0.0138 & 0.1967 \\

\hline
\hline
\end{tabular}
\caption{{\bf Samples and structural parameters.} Volume fractions $\phi$ of the samples used in this work, together with the nearest-neighbour peak position, $q_1$, corresponding average particle separation, $d$, and first peak half-width, $w_1$. Also shown is the relative change of the first peak position, $\Delta q_1/q_1$, at an applied shear rate of $10^{-1} s^{-1}$.}
\label{tab:phi}
\end{table}

Anisotropic structure factors have been observed in strongly sheared colloidal liquids as continuous distortions from the liquid equilibrium structure~\cite{Ackerson1991,Cohen2011}. However, for out-of-equilibrium supercooled liquids and glasses, such changes are small and difficult to measure. Some insight into the observed anisotropy comes from direct real-space imaging of particle rearrangements in glasses~\cite{chikkadi_schall11,schall2007}: the shear component of the microscopic strain field shows a quadrupolar symmetry that reflects the elastic response of the material to local shear transformations~\cite{schall2007,picard2004}. The normal strain component of this distortion field~\cite{Eshelby} has a $p$-wave symmetry in the shear plane, which is precisely the symmetry that we observe here. In the ideal case of $p$-wave symmetry, however, the two maxima of the correlation function $C(\beta)$ should be equal exactly to 1. In contrast, the experimental picture we observe here is more complex - the value of the second maximum $C(\beta=\pi)$ is less than 1 during shear and shortly after, indicating interesting intermediate states with orientational correlations in the structure of the relaxing glass. Such intermediate states are associated with systems that are dynamically out of equilibrium~\cite{Struth2011}, and exhibit memory of their history, such as for example glassy and amorphous systems.
This is again reflected in the relaxation of the glass (pink dashed line in Fig. 2b), which occurs on a time scale close to $t_{s}$, similar to Fig. 1b.

These results are important for constitutive models of the deformation of amorphous materials, where the effect of shear is described by a structural parameter, often referred to as "effective temperature" \cite{Langer,Bouchbinder}. An important prediction of these theories is the occurrence of shear banding instabilities that result from a local coupling between shear and structure. While direct real-space imaging~\cite{chikkadi_schall11} clearly establishes the occurrence of shear banding at shear rates of the order of the inverse structural relaxation time, these measurements cannot resolve the small structural change of the glass upon shear banding, and the coupling between shear and structure remains unresolved. Using the combined x-ray and rheology measurement we obtain new insight into this coupling. We varied the applied shear rate by a few orders of magnitude around the inverse structural relaxation time, $t_s^{-1} \sim 3 \times 10^{-2}s^{-1}$, where we observed the emergence of shear banding by microscopy~\cite{chikkadi_schall11}.

Indeed, our rheological measurements (Fig. 3a) show the characteristic plateau of a shear banding instability at $\tau \sim 140$ Pa (arrow) \cite{SB}. This plateau indicates that at the same applied stress the two strain rates $10^{-3}$ s$^{-1}$ and $3 \times 10^{-2}$ s$^{-1}$ coexist. %Repeated measurements on a suspension with a slightly lower volume fraction showed a similar plateau regime, but shorter and shifted to higher shear rate as expected, thus lending further credence to our interpretation of shear banding.
Having thus established the shear-banding regime, we can now look at the steady-state structures, measured at the same time (Fig. 3b). Interesting non-monotonic behavior is observed in all structural parameters, indicating a structural transition that accompanies the shear banding. To investigate the coexistence of structural states, we plot two representative structural parameters, $\Delta q_1$ and $\Delta w_1$ as a function of stress in the inset of Fig. 3a. Indeed, at the plateau stress $\tau \sim 140$ Pa, different values of these parameters, positive and negative, coexist suggesting the coexistence of two structural states. We further confirm the existence of shear banding by varying the separation of the two rheometer plates: when we decreased the plate separation, we observed much weaker non-monotonous behavior, as expected.

\begin{figure}
\centering
\subfigure
{\includegraphics[width=0.47\columnwidth]{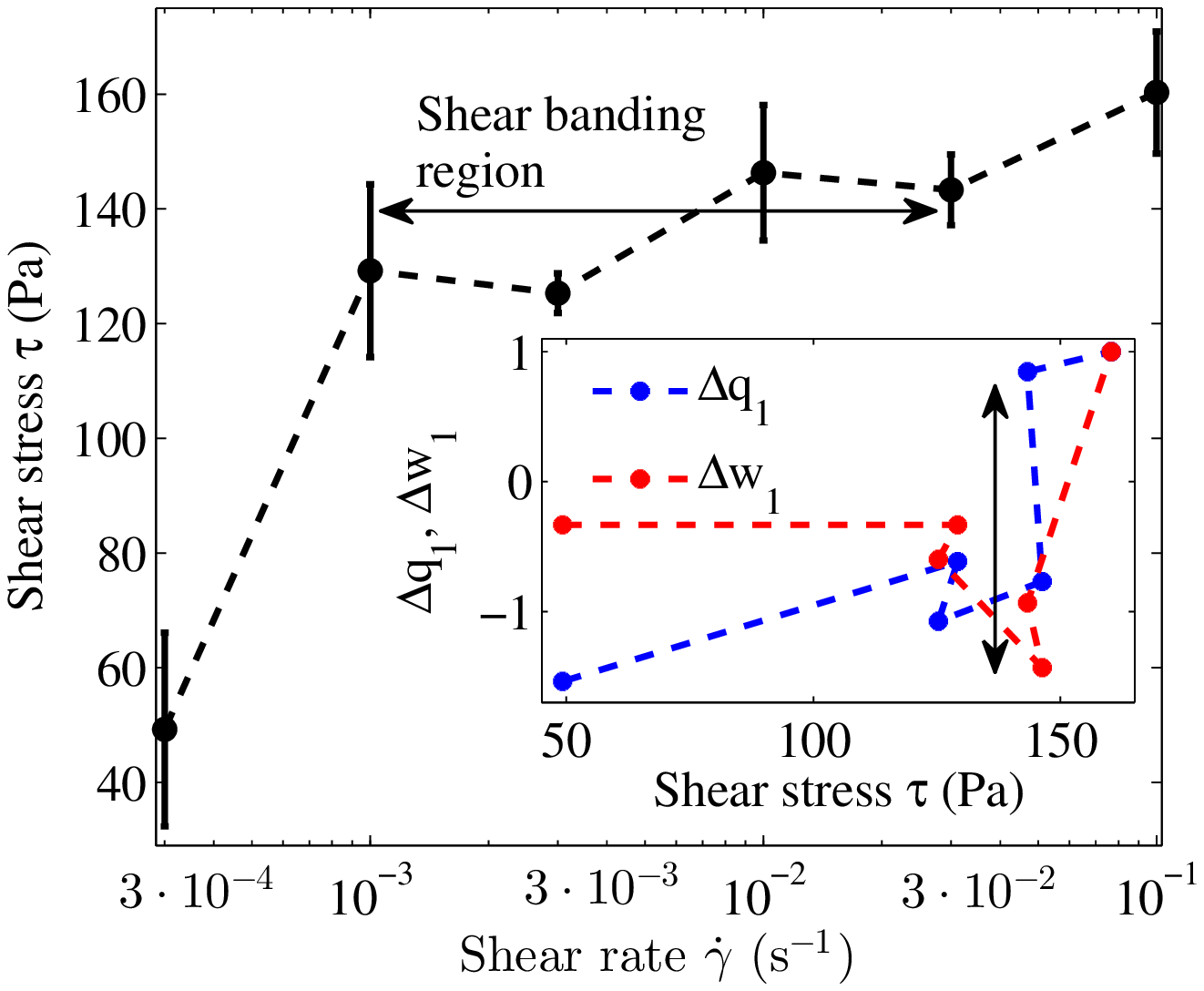}
\label{fig:StressVsShear}
}
\begin{picture}(0,0)(0,0)
\put(-200,00){(a)}
\end{picture}
\subfigure
{\includegraphics[width=0.43\columnwidth]{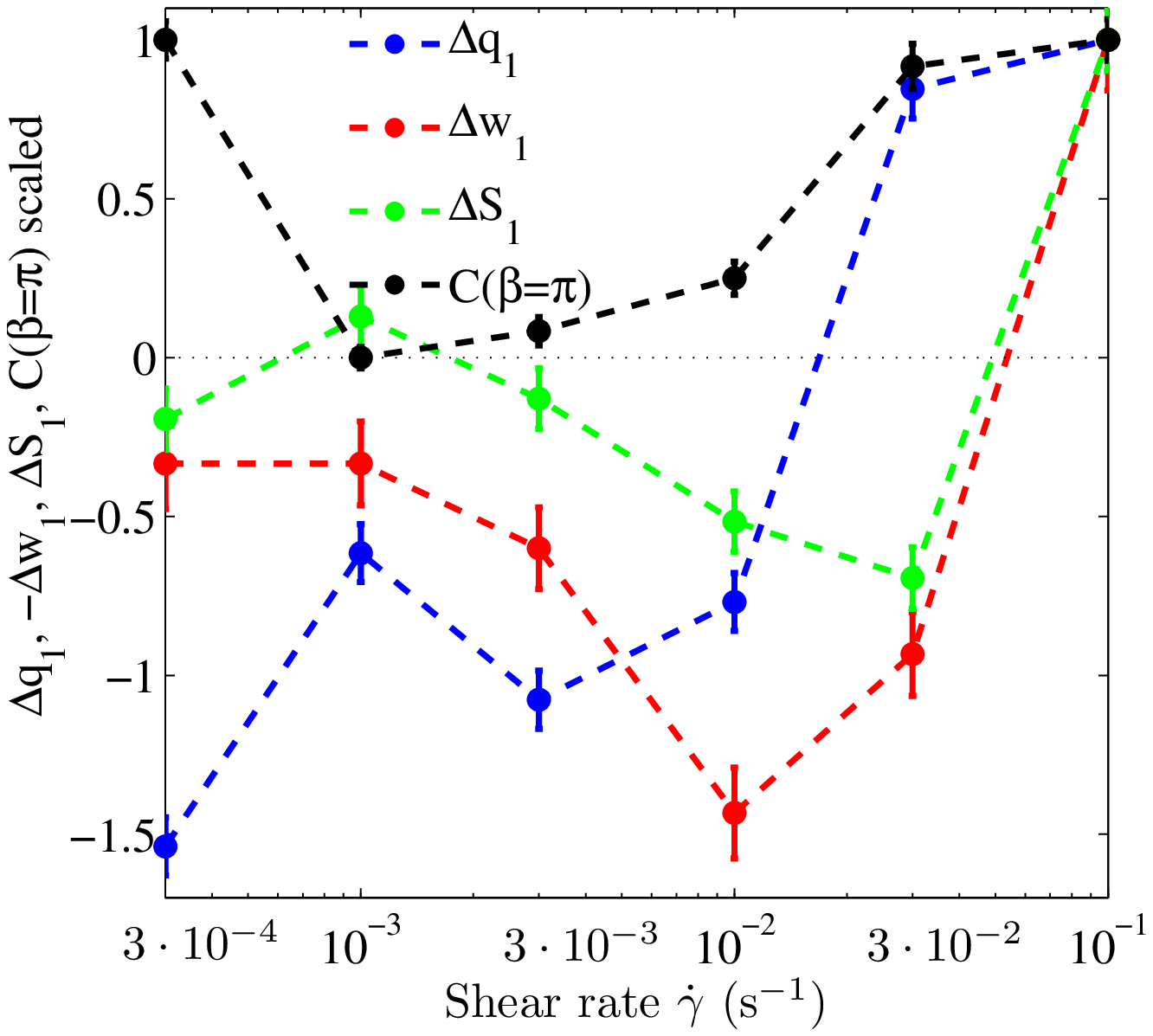}
\label{fig:structure_change}
}
\begin{picture}(0,0)(0,0)
\put(-200,00){(b)}
\end{picture}
\caption{
{\bf Shear-Structure coupling at the shear banding transition.}
(a) Flow curve indicates a shear banding regime between $\dot{\gamma} = 10^{-3}$ and $3 \cdot 10^{-2}$ s$^{-1}$ (arrow), vertical lines show the error margins; measurements are limited to six different strain rates due to time limitations. Inset: $\Delta q_1$ and $\Delta w_1$ as a function of shear stress reveal the coexisting structures of the shear bands (arrow).
(b) Structural parameters as a function of shear rate $\dot{\gamma}$ for the volume fraction $\phi = 58\%$. Positive values of $C(\beta = \pi)$ (black line) indicate appearance of two-fold symmetry. All quantities are scaled to unity. Least-square fits have been used to determine the average peak position during shear and relaxation. Vertical lines show the corresponding error margins.
}
\end{figure}

\section*{Discussion}

The similar non-monotonic variation of structural parameters and flow stress suggests that there is significant coupling between shear and structure. The onset of shear banding is clearly manifested in the appearance of a structurally liquid-like state: the structural anisotropy characteristic of the solid disappears as indicated by the vanishing value of $C(\beta)$ for $10^{-3}\leq\dot{\gamma}\leq10^{-2}$ s$^{-1}$ (Fig. 3b). At the same time particle separations increase, and fluctuations become more enhanced as reflected in the negative values of $\Delta q$ and $\Delta w_1$, again indicating a liquid-like state. Such liquid-like response is known from the shear banding of metallic glasses that show characteristic liquid vein patterns along the shear band~\cite{spaepen_75}. Our direct particle tracking by microscopy has indeed revealed liquid-like mean square displacements in the shear band~\cite{chikkadi_schall11}; nevertheless, no change in the structure could be detected. The combined x-ray and rheology measurements presented here allow us to resolve the distinct structural transition that underlies the shear banding instability.

While the structures of the coexisting bands cannot be easily resolved separately in the current geometry, we can separate the structural properties of the bands by assuming that the portions of the two bands follow a lever rule, and their structural properties remain unchanged~\cite{Moller2008}. We can then distinguish the coexisting structures by plotting the structural parameters as a function of stress, as we have shown in Fig. 3a inset; the two values of $\Delta q_1$ and $\Delta w_1$ reflect the coexisting structures of the two bands. The positive and negative changes of $\Delta q$ and $\Delta w_1$ correspond to $0.5\%$ and $3\%$ of $\Delta q$ and $\Delta w_1$, respectively, small but distinct changes that give rise to largely different rheological properties.

By using a new combination of rheology and x-ray scattering, we have demonstrated the distinct structural transition that a glass undergoes under applied shear. This combination allowed us for the first time to establish an immediate connection between applied stress and structure in the genuine shear banding instability of glasses. Our results reveal a coupling between structural parameters and the applied shear that underlies this instability: the non-monotonic behavior of the flow curve is directly mirrored in simple structural measures such as the position, the width, and the height of the nearest neighbor peak. Besides small changes in the nearest neighbor distances, our results underscore the importance of anisotropy in the structure of such out-of-equilibrium systems, in agreement with recent structure analysis of jammed and unjammed granular packings~\cite{SchroderTurk2010}. These measurements provide important test cases for constitutive models and mode coupling predictions of the deformation of amorphous materials.

\section*{Methods}
We use silica particles with a diameter of $50$ nm and a polydispersity of $10\%$, suspended in water. A small amount of $NaCl$ is added to screen the particle charges. We estimate the Debye screening length to be $2.7$ nm, resulting in an effective diameter of $2r_0 = 55.4$ nm. Samples with effective volume fractions of $\phi=34.5\%$, $58\%$ and $63.5\%$ ($\pm2\%$) are prepared by diluting the sediment after centrifugation (see Table~\ref{tab:phi}). A rheometer in plate-plate geometry is filled with the suspension. To image the suspension under shear, the well-collimated synchrotron beam with a wavelength of $\lambda = 0.154$ nm is deflected in the vertical direction, and directed through the layer of suspension perpendicular to the rheometer plates, see Fig. 1a. The scattered intensity is measured at a distance of $D = 313$ cm using a Pilatus detector with a pixel size of $172 \times 172$ $\mu$m$^2$, covering scattering angles $\theta$ in the range $0.03-0.5^{\circ}$. This allows us to resolve structural changes in the shear plane for wave vectors $q = 4\pi/\lambda sin(\theta/2)$ in the range $qr_0 = 0.5$ to $5$. Before each measurement, the samples are initialized by preshearing at a rate of $\dot{\gamma}=0.1$ s$^{-1}$ for 120 seconds, and then left to rest for 600 seconds to obtain reproducible results. We then shear the suspensions at rates between $\dot{\gamma} = 10^{-4}$ and $10^{-1}$ s$^{-1}$ well into the steady state, followed by $600$ seconds of relaxation. The scattered intensity is recorded at a frame rate of 1 s$^{-1}$ during shear and relaxation. We determine the structure factor $S(\textbf{q})$ from the recorded intensity by subtracting the solvent background and dividing by the form factor, determined from dilute suspensions.

\section*{Acknowledgements}

We thank Matthias Fuchs for helpful discussions. We thank Truc Anh Nguyen for help with the experiments. The authors thank DESY, Petra III for access to the X-ray beam. This work was supported by the Foundation for Fundamental Research on Matter (FOM) which is subsidized by the Netherlands Organisation for Scientific Research (NWO). P.S. acknowledges support by a Vidi fellowship from NWO.

\end{document}